# HIGH THROUGHPUT OPEN-SOURCE IMPLEMENTATION OF Wi-Fi 6 AND WiMAX LDPC ENCODER AND DECODER

Tomáš Páleník, Viktor Szitkey

*Abstract:*

*This paper describes the design and C99 implementation of a free and open-source Low-Density Parity-Check (LDPC) codes encoder and decoder focused primarily on the Quasi-Cyclic LDPC (QC-LDPC) codes utilized in the IEEE 802.11ax-2021 (Wi-Fi 6) and IEEE 802.16-2017 (WiMAX) standards. The encoder is designed in two variants: the first one universal, the other a minimal memory usage design. The decoder provides a single- and multi- threaded implementation of the layered single-scan min-sum LDPC decoding algorithm both for floating point and fixed-point arithmetic. Both encoder and decoder are directly callable from MATLAB using the provided MEX wrappers but are designed to be simply used in any C project. A comparison of throughput and error performance with the recent commercial closed-source MEX implementation of an LDPC encoder and decoder introduced in MATLAB R2021b Communications Toolbox is provided. Source code portability to alternative non-x86 architectures is facilitated by using only the standard C99 constructs, GNU tools, and POSIX libraries. The implementation maintains low-memory requirements, enabling its deployment in a constrained-architecture in the context of Internet of Things. All source codes are freely available on GitHub under a permissive BSD license.*

*ACM Computing Classification System:*

*Mathematics of computing, Mathematical software, Mathematical software performance.*

## Introduction

After the introduction of the QC-LDPC codes in [1] and the desired direct-encoding [2] calculated using the sparse parity check matrix **H** without needing the code generator matrix G, QC-LDPC codes have been an integral part of modern communication standards for some time now. This ranges from IEEE 802.16 [3], through IEEE 802.11ax [4,5], DVB-S2 [6] up to the latest 5G 3GPP Release 17 TS 38.212 [7], and even the wired networking Ethernet IEEE 802.3 standard [8]. Because of their faster decoding, the ultra-high speed 200 Gbps and 400 Gbps IEEE 802.3-2018 prescribe the use of Reed Solomon (RS) codes [9] to fit the strict latency constraints. Even though the RS codes don't reach the exceptional near-capacity-limit error performance of LDPC, the research in this complementary area is still ongoing [10]. Arguably the most sophisticated LDPC designs are presented in the CCSDS specifications, such as [11] and [12], with intense research activity ongoing: Recent papers [13], [14], [15], propose novel hardware encoders, for the CCSDS standard QC-LDPC codes, achieving acceleration primarily by restructuring encoder structure and utilizing parallelism. [16] provides a comparison of several hard- and soft-decision decoding algorithms for the CCSDS defined (128512, 64256) QC-LDPC code, while [17] deals with potential energy saving in a satellite link. [18] describes how the Automorphism Ensemble Decoding (AED) can be enabled for QC-LDPC codes to improve error performance for codes with shorter $N$ (128 to 256) used in Wi-Fi, 5G, and CCSDS specifications and [19] compares the Belief propagation, Min-Sum and Neural Normalized Min-Sum (N-NMS) decoding performance in the context of line product codes (LPC) also defined by the CCSDS. [20] deals with modifying existing LDPC codes to incorporate desired run length properties and [21] presents a mathematical framework for constructing QC-LDPC codes with desired girth, providing examples based on the CCSDS-defined protographs.

In any case, the LDPC codes have evolved to become ubiquitous in almost all broadband wireless technologies. Implementations of encoders and decoders are also plentiful [22] - [26], with [27] giving an exhaustive overview of even more. These are all, however, hardware implementations: either ASIC designs intended to be integrated into modems, or FPGA-based. Complementing these are pure software implementations available at GitHub: [28] provides universal encoder/decoder C++ implementations, but doesn't include the practical, standard-specified **H** matrix designs. [29] provides the IEEE802.11 codes and separate MATLAB and C encoder/decoder implementations. [30] focuses on CCSDS LDPC codes and provides a MATLAB implementation. In [31] a C++ implementation defines a SIMD-acceleration using the AVX-512 vectorized arithmetic focused on the DVB-S2 and DVB-T2 standard. [32] provides a MATLAB implementation for the 5G New Radio TS38.212 encoder and decoder. [33] is an older C implementation focused on education with no standard **H** matrix designs, and [34] describes the results for a SIMD-accelerated x86-specific C++ encoder/decoder implementation using the Intel compiler and the optimized Math Kernel Library with focus on the CCSDS and DVB-S2 LDPC codes.

The common denominator of all these implementations is the lack of integration of fast C/C++ code (if available) with MATLAB. This integration is important, since it is impractical to write whole simulations in C or maintain two parallel (and potentially functionally different) versions for MATLAB and C/C++. The obvious solution would be the utilization of fast C/C++ code in MATLAB by implementing MEX-file wrappers that allow calling such a method directly from MATLAB. Such implementation exists as a commercial closed-source product and is part of the Communications Toolbox. In fact, the LDCP codes in MATLAB have a history on their own: the first introduction of LDPC in MATLAB came in 2007 in the form of the `dvbs2ldpc()` function, still available today. In 2012 the `comm.LDPCDecoder` and `comm.gpu.LDPCDecoder` system objects were introduced in R2012a [35], with the former now deprecated and replaced by the universal `ldpcDecode()` function, complemented by `ldpcEncode()`, both being new improved and optimized implementations introduced only recently in R2021b [36]. These come with the ability to expand practical QC-LDPC prototype matrices to sparse **H** matrices, but the task of obtaining the prototype matrices is left to the user. The implementations are available as binary MEX modules only, so it is hard to infer what advanced optimizations were made. Furthermore, for a researcher, who would like to test potential modifications to the encoding/decoding algorithms, while evaluating practical LDPC codes used in modern standards, such closed-source implementations are of little use.

This is where our implementation comes to play: it provides a relatively fast C99 implementation of a QC-LDPC encoder and decoder that can be integrated into any C project, directly containing practical LDPC codes utilized in the IEEE802.11ax-2021 [4] (actually defined in the earlier IEEE802.11-2020 specification [5]) and IEEE 802.16-2017 standards, while also providing MEX wrappers and supporting functions for seamless MATLAB integration for statistical evaluation of waterfall curves. Everything is available as completely free and open source code under the permissive BSD license. To facilitate portability, and even a potential ARM platform compatibility, no external libraries, such as the optimized Intel MKL, or ISA-specific constructs, such as intrinsics, were used.

The rest of the paper is organized as follows: the next section contains a brief recap of the QC-LDPC codes design along with some important specific parameters of the practical codes used in IEEE802.11ax and IEEE802.16-2017. The second section recaps the direct encoding algorithm and the fourth describes the two encoder implementations: one universal and another designed with minimal memory requirements in mind. These can be switched at compile-time. The next section then recaps the design and describes our implementation of the layered single-scan min-sum decoder algorithm, while the sixth section provides some thoughts on a fixed point arithmetic decoder design. The sixth section provides an error performance evaluation in the form of waterfall curves, along with the comparison of throughput of various decoder setups in contrast to the existing MATLAB Communications Toolbox implementation. The last section concludes the paper.

# 1 LDPC Codes in Modern Standards

Discovered by Gallager [37], the LDPC codes are linear block codes (LBC) with basic parameters [ $N$, $K$ ] defined by their sparse parity check matrix **H**. The code generator matrix $\mathbf{G}^{K \times N}$ can be derived from $\mathbf{H}^{M \times N}$ (defining: $M = N - K$ and code rate $R = K / N$), but it will unlikely be sparse and therefore unsuitable for practical high-throughput encoding. The encoding using the **G** matrix is, however, described very simply by multiplication of the data word – the vector **i** of length $K$ bits by **G** to obtain the codeword **c** of length $N$. For a systematic code, the systematic part **s** of the codeword is a subvector of **c** equal to **i**. We may denote:

$$\mathbf{c}^{1 \times N} = \begin{bmatrix} \mathbf{s}^{1 \times K} & \mathbf{p}^{1 \times M} \end{bmatrix} = \mathbf{i}^{1 \times K} \cdot \mathbf{G}^{K \times N}, \mathbf{s} = \mathbf{i} \qquad (1)$$

The sparsity of the large **H** matrix not only provides for the excellent error correction properties of LDPC codes, it also allows for an efficient storage of **H** in memory: instead of using an array of size $M \times N$, only the indices of ones need to be stored. Throughout the rest of the paper we will use $n$ for the column index (also known as the variable node index in soft-decoding context) and $m$ for the row index of **H** (also known as the check-node index). Then for each $n \in \{ 1, 2, \ldots, N \}$ the set of row indices of ones in a given column of **H** may be called neighborhood of $n$ and denoted $M(n)$ while for each $m \in \{ 1, 2, \ldots, M \}$ the $N(m)$ represents the set of column indices of ones in a given row $m$. Any sparse binary matrix **H** can be stored in a memory efficient way as an array of arrays $M(n)$, or array of arrays $N(m)$. The number of elements of $g(m) = |N(m)|$ may be called the degree of row $m$, while the symbol $f(n) = |M(n)|$ denotes the degree of column $n$. For practical irregular LDPC codes, the $g(m)$ and $f(n)$ are not constant over $m$ and $n$, and their average and maximum values are of interest.

$$\mathbf{H} = \begin{bmatrix} \mathbf{P}_{0,0} & \mathbf{P}_{0,1} & \cdots & \mathbf{P}_{0,N_b-1} \\ \mathbf{P}_{1,0} & \mathbf{P}_{1,1} & & \mathbf{P}_{1,N_b-1} \\ \vdots & & & \\ \mathbf{P}_{M_b-1,0} & \mathbf{P}_{M_b-1,1} & \cdots & \mathbf{P}_{M_b-1,N_b-1} \end{bmatrix} \qquad (2)$$

The subclass of practical QC-LDPC codes used in communications standards [3-5] was deliberately constructed with such structure of the **H** matrix that allows for the encoding process to be completely and efficiently implemented directly using **H**. As shown in (2), the very large binary sparse matrix **H** is designed as a block matrix consisting of a grid of submatrices $\mathbf{P}_{i,j}$ of size $Z \times Z$.

Each of the submatrices may be a zero matrix or a permuted identity matrix, with the only allowed permutation being a cyclic shift, resulting with $\mathbf{P}_{i,j}$ circulant and sometimes called a cyclic shift matrix.

The constraints on $\mathbf{P}_{i,j}$ allow for a compressed representation of $\mathbf{H}$ in a form of a model matrix $\mathbf{H}_{bm}$ of size $M_b \times N_b$ (denoting $M_b = N_b - K_b$) where each element $\mathbf{H}_{bm}(i, j)$ represents the circulant permutation matrix $\mathbf{P}_{i,j}$ and $N = N_b \times Z$, $K = K_b \times Z$. By convention the value of $\mathbf{H}_{bm}(i, j)$ equal to -1 represent a zero submatrix $\mathbf{P}_{i,j}$, value 0 an identity submatrix, and any positive number represents identity matrix circularly shifted by $\mathbf{H}_{bm}(i, j)$. The compressed $\mathbf{H}_{bm}$ matrices are then directly specified by the standards with different model matrices defined for each codeword length $N$, as in [5], or common structure for the largest value of $N$ given along with a corresponding transformation for all other supported values of $N$, as in [3]. In all cases the large sparse binary matrix $\mathbf{H}$ is obtained from $\mathbf{H}_{bm}$ by expanding each element of $\mathbf{H}_{bm}$ to a square matrix of size $Z$. For both standards of interest, the $\mathbf{H}$ matrices share further similarity: The $\mathbf{H}_{bm}$ can be further partitioned to three submatrices as follows (using the notation from [3]):

$$\mathbf{H}_{bm}^{M_b \times N_b} = \left[ \mathbf{H}_{b1}^{M_b \times K_b} \quad \mathbf{H}_{b2}^{M_b \times M_b} \right] = \left[ \mathbf{H}_{b1}^{M_b \times K_b} \quad \mathbf{h}_{b}^{M_b \times 1} \quad \mathbf{H}_{b2'}^{M_b \times (K_b - 1)} \right] \quad (3)$$

In terms of submatrix indexing the submatrices $\mathbf{H}_{b1}$, $\mathbf{h}_b$ and $\mathbf{H}_{b2'}$ may be defined as:

$$\mathbf{H}_{b1} = \mathbf{H}_{bm}[0, 1, ..., M_b - 1; 0, 1, ..., K_b - 1] \quad (4)$$

$$\mathbf{h}_{b} = \mathbf{H}_{bm}[0, 1, ..., M_b - 1; K_b] \quad (5)$$

$$\mathbf{H}_{b2'} = \mathbf{H}_{bm}[0, 1, ..., M_b - 1; K_b + 1, ..., N_b - 1] \quad (6)$$

where $\mathbf{h}_b$ is the first column of matrix $\mathbf{H}_{b2}$. We use the a slightly mathematically imprecise indexing starting at zero instead of one. We do this in order to be compatible with the indexing defined in the standard [3].

The direct encoding algorithm further requires that the column vector $\mathbf{h}_b$ has a special structure containing two paired (equal) values in the first and last position and a single positive value among the remaining positions. All other elements are set to $-1$ indicating later expansion to zero matrices. This structure is required for the direct encoding described in later sections. Also required is the structure of the submatrix $\mathbf{H}_{b2'}$, where two zero diagonals within an all $-1$ matrix in $\mathbf{H}_{b2'}$ translate to double diagonal expanded submatrix of $\mathbf{H}$. Equation (7) shows the example structure of $\mathbf{h}_b$ and $\mathbf{H}_{b2'}$ for $N_b = 24$, $K_b = 18$ for the $R = 3/4$ QC-LDPC code defined in [3]:

$$\mathbf{H}_{b2}^{6 \times 6} = \left[ \mathbf{h}_{b}^{6 \times 1} \quad \mathbf{H}_{b2'}^{6 \times 5} \right] = \begin{bmatrix} \begin{bmatrix} 0 \\ -1 \\ 80 \\ -1 \\ -1 \\ 0 \end{bmatrix} & \begin{bmatrix} 0 & -1 & -1 & -1 & -1 \\ 0 & 0 & -1 & -1 & -1 \\ -1 & 0 & 0 & -1 & -1 \\ -1 & -1 & 0 & 0 & -1 \\ -1 & -1 & -1 & 0 & 0 \\ -1 & -1 & -1 & -1 & 0 \end{bmatrix} \end{bmatrix} \quad (7)$$

Both standards specify a set of supported code rates and codeword sizes that further influence the design of encoder and decoder implementation. Parameters for the IEEE 802.11ax standard are summarized in (Tab.1), while the IEEE802.16-2017 is summarized in (Tab.2).

Table 1. Data word size *K*, codeword size *N*, and expansion factor *Z* for the IEEE 802.11ax.
Word sizes divisible by 8 bits are shown in bold.

| *N*(bit) | *Z*(bit) | *K* (bit) | | | |
|---|---|---|---|---|---|
| | | *R*=1/2 | *R*=2/3 | *R*=3/4 | *R*=4/5 |
| 648 | 27 | 324 | **432** | 486 | 540 |
| 1296 | 54 | **648** | **864** | 972 | **1080** |
| 1944 | 81 | 972 | **1296** | 1458 | 1620 |

It is important to note that different sections of the Wi-Fi 6 standard define several different LDPC codes and that different versions of each standard may share some LDPC codes. In this section we will deal with the codes defined in the IEEE 802.11-2020 [5], since these provide the longest codeword sizes and best error correcting capabilities. These are then also reused in the more recent standard version IEEE 802.11ax-2021 [4].

As show in (Tab.1), all codeword lengths *N* are multiples of 8 bits. This is not true for all data word sizes *K*. The expansion factor (or subblock size) *Z* is deliberately chosen to take on values 27, 54 or 81, clearly with the implementation of a hardware encoder in the form of specifically designed ASIC circuit in mind. As discussed in later sections, this design choice will complicate a memory optimized C implementation. In terms of alignment of bits to blocks easily fitting standard C data types, a slightly better situation is with the IEEE802.16-2017 standard, shown in (Tab.2), where all data words and all codewords fit the 8bit-in-a-byte storage. Also depicted are all the possible values of the subblock size *Z*. Starting at 24 and incrementing by 4 up to 96, 10 out of the total 19 values are divisible by 8, facilitating an optimized binary encoder implementation. Values 32, 64, and 96 enable further optimization by using wider native C types.

Table 2. Data word size *K*, codeword size *N*, and subblock size *Z* for the IEEE 802.16-2017 standard.
All data and codewords lengths in bits are multiples of 8. Only half of the *Z*s are divisible by 8,
while all data word sizes fit a 8bit-in-a-byte storage. *Z* factors divisible by 8 are show in bold.

| *N*(bit) | *Z*(bit) | *K* (Byte) | | | |
|---|---|---|---|---|---|
| | | *R*=1/2 | *R*=2/3 | *R*=3/4 | *R*=4/5 |
| 576 | **24** | 36 | 48 | 54 | 60 |
| 672 | 28 | 42 | 56 | 63 | 70 |
| 768 | **32** | 48 | 64 | 72 | 80 |
| 864 | 36 | 54 | 72 | 81 | 90 |
| 960 | **40** | 60 | 80 | 90 | 100 |
| 1056 | 44 | 66 | 88 | 99 | 110 |
| 1152 | **48** | 72 | 96 | 108 | 120 |
| 1248 | 52 | 78 | 104 | 117 | 130 |
| 1344 | **56** | 84 | 112 | 126 | 140 |
| 1440 | 60 | 90 | 120 | 135 | 150 |
| 1536 | **64** | 96 | 128 | 144 | 160 |
| 1632 | 68 | 102 | 136 | 153 | 170 |
| 1728 | **72** | 108 | 144 | 162 | 180 |
| 1824 | 76 | 114 | 152 | 171 | 190 |
| 1920 | **80** | 120 | 160 | 180 | 200 |
| 2016 | 84 | 126 | 168 | 189 | 210 |
| 2112 | **88** | 132 | 176 | 198 | 220 |
| 2208 | 92 | 138 | 184 | 207 | 230 |
| 2304 | **96** | 144 | 192 | 216 | 240 |

## 2 QC-LDPC Direct Encoding

The special structure of the **H** matrix was designed to allow for the direct encoding algorithm where the encoding of a systematic code uses the **H** matrix instead of the code generator matrix **G**. For a practical C encoder implementation the model matrix **H**$_{bm}$ is actually used, so the large sparse binary **H** matrix, or even permutation submatrices **P**$_{i,j}$ are never actually stored anywhere in memory. Since the multiplication of some subvectors of **i** of length $Z$ by matrix **P**$_{i,j}$ is identical in result to cyclic shift, the actual operation to implement as a building block of the encoder is the cyclic shift of subvectors of size $Z$ bits by **H**$_{bm}(i, j)$ positions. The direct encoder algorithm originally introduced in [38] and described in sufficient detail also in [3] can be summarized as follows:

The information block **i** (also known as systematic vector s) is divided into $K_b$ blocks of $Z$ bits. The grouped **i** can be denoted as a matrix **u**, where each element $u_i$ is a column vector:

$$\mathbf{u}^{Z \times K_b} = [u_0 \quad u_1 \quad \ldots \quad u_{K_b-1}] \tag{8}$$

$$u_i = [s_{iZ} \quad s_{iZ+1} \quad \ldots \quad s_{(i+1)Z-1}]^T \tag{9}$$

The parity sequence **p** is calculated in blocks of $Z$ bits directly using the matrix **H**$_{bm}$. Again, the parity sequence **p** can be expressed as a matrix in terms of blocks, where each element $v_i$ is a column vector $Z$ bits long:

$$\mathbf{p}^{Z \times M_b} = [v_0 \quad v_1 \quad \ldots \quad v_{M_b-1}] \tag{10}$$

$$v_i = [p_{iZ} \quad p_{iZ+1} \quad \ldots \quad p_{(i+1)Z-1}]^T \tag{11}$$

The slightly nonstandard notation for vectors $u_i$ and $v_i$ is used here in order to preserve notation in [3]. The encoding is divided into initialization and recursion. First the initialization is defined as follows:

$$v_0 = \mathbf{P}^{-1}_{\mathbf{H}_{bm}(y,K_b)} \cdot \left( \sum_{i=0}^{M_b-1} \sum_{j=0}^{K_b-1} \mathbf{P}_{\mathbf{H}_{bm}(i,j)} u_j \right) \tag{12}$$

Where **H**$_{bm}(i, j)$ is an element of the (potentially scaled) model matrix **H**$_{bm}$, **P**$_{\mathbf{H}bm(i, j)}$ represents the circulant permutation matrix, and multiplication by **P**$_{\mathbf{H}bm(i, j)}$ implements the cyclic shift operation on subvector $u_j$. The element **H**$_{bm}(y, K_b)$ is the only nonnegative element of the subvector **h**$_b$ outside of the paired values, as shown in the example in (7) with **H**$_{bm}(y, K_b) = $ **h**$_b(2) = 80$. Note that the value of $y$ is different for each **H**$_{bm}$ and must be found on-the-fly by the encoder. The inverse permutation **P**$^{-1}$ is just a cyclic shift in the opposite direction. All addition/sum operations are performed over GF(2).

After initialization and successful calculation of $v_0$, the encoder proceeds with iteration/recursion, finding the remaining parity subblocks $v_i$:

$$v_1 = \mathbf{P}_{\mathbf{H}_{bm}(0,K_b)} v_0 + \sum_{j=0}^{K_b-1} \mathbf{P}_{\mathbf{H}_{bm}(0,j)} u_j \;,\; i = 0 \tag{13}$$

$$v_{i+1} = v_i + \mathbf{P}_{\mathbf{H}_{bm}(i,K_b)} v_0 + \sum_{j=0}^{K_b-1} \mathbf{P}_{\mathbf{H}_{bm}(i,j)} u_j \;,\; i > 0 \tag{14}$$

Since the sums terms in (12):

$$\mathbf{S}_i \equiv \sum_{j=0}^{K_b-1} \mathbf{P}_{\mathbf{H}_{\mathbf{bm}}(i,j)} u_j \ , i = 0 ... M_b - 1 \tag{15}$$

are calculated during the initialization and reused during the iteration, it is advantageous to store their values for reuse. Since the elements of $\mathbf{H_{bm}}$ equal to -1 expand to all zero $\mathbf{P}$ submatrix, a multiplication by such a $\mathbf{P_{Hbm}}$ results to a zero term in sums and may be safely skipped in the implementation.

## 3 C99 Encoder Implementation

The mathematical definition of the encoder given in the previous section determines the design of the final C algorithm, while the actual values of code parameters *N*, *K*, *M*, and *Z* play an important role. This paper describes two encoder implementations. The first one is a universal array implementation, that supports all combinations of the code parameters, but requires more memory, since each bit is actually stored as a C array element (usually but not necessarily an 8bit-wide `unsigned char`). The advantage is a simple code, where bit accesses are directly supported by an array access operator. The disadvantage for actual transmission systems/modems processing blocks of truly binary data is the necessity to expand the data inside the encoder by a factor of 8, so that each bit is actually stored in a separate byte.

The second implementation is a bitmap (or packed-bits) encoder – a true binary encoder in that it handles binary data blocks as bitmaps, and accesses bits within bytes by using the bitwise operators. While it is possible to implement cyclic shifts even for a subblock of size *Z* defined in the Wi-Fi standard {27, 54, 81}, these require non-byte aligned memory accesses that hurt performance. This approach was tried and benchmarked, and the measured low throughput was then a reason for not implementing the bitmap encoder for these values of *Z*. On the other hand, the Wi-MAX standard provides several code parameters combinations, where all *N*, *K*, *M*, *Z* are divisible by 8 (or by 16 and even 32 and 64) which allows for a clean and fast C implementation with minimal memory requirements.

All encoder equations (12) - (15) were implemented directly, replacing the multiplication $\mathbf{H_{bm}}(i, j).\ u_j$ with a cyclic shift of vector $u_j$ by $\mathbf{H_{bm}}(i, j)$ bit positions and each addition with a bitwise XOR. The C language doesn't directly support a construct that would access the underlying ISA rotation instructions, so two linear shifts of a variable are called and combined with a bitwise OR operator. This is one of the known drawbacks of the C language [39] and the task of compiling the three related C operations into one rotation instruction is left to the compiler.

Since the encoding algorithm is defined in terms of blocks of bits of size *Z*, that can span several bytes or multibyte words, array indexing and bitwise operators must be combined to implement the circular shift that is the basis of equations (12) - (15). The principle of the memory efficient bitmap encoder cyclic shift routine is demonstrated in (Fig.1), where three indices are defined: the bit index $I_{bBL}$ within the block of size *Z*, the bit index within the word $I_{bW}$ and the index $I_W$ of the word within the array. Since the array is stored in memory addressed in bytes, the byte index $I_B$ denotes the byte offset from the start of the array. It may be a convenient, but not necessary to set the word size to contain just one byte, making $I_W = I_B$. This choice is made in (Fig.1) for the sake of example simplicity. Let *wb* denote the word size in bits.

Also depicted in (Fig.1), is the observation that the shift *s* can be conceptually divided into two parts: the shift *sw* expressed in an integral number of words and the remaining shift of up to *wb* – 1 bits. Shown for selected values of *s* = 3, 11, 19, all congruent modulo *wb* = 8, the resulting array elements, although placed on different array positions, are bitwise identical ( [ ] denotes C array indexing): $Out_s=11[1] = Out_s=3[0] = Out_s=19[2]$.

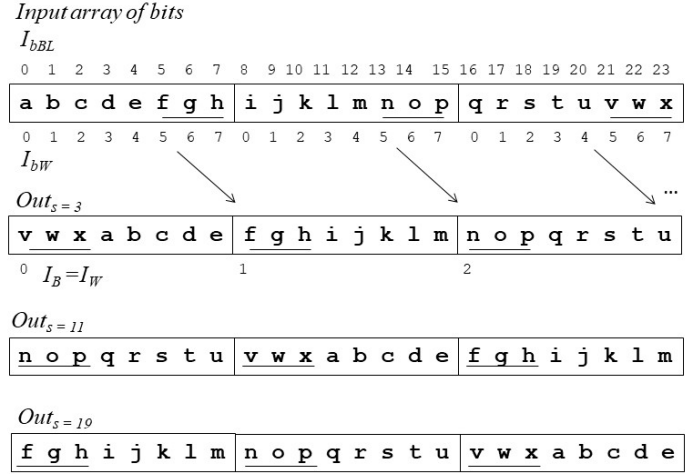

Fig.1. The array cyclic shift examples for word size $wb = 8$ bits, block size $Z = 24$ and selected shift value $s = 3; 11; 19$. Indices are: index of the bit within the block $I_{bBL}$, of the bit within the word $I_{bW}$, of the word $I_W$ in the array, and the byte offset from the start of the array $I_B$ (now same as $I_W$)).

The arbitrary cyclic shift by $s$ positions can be represented as $s = sw \cdot wb + sb$ where $sw = \text{div}(s, wb)$ is the partial rotation of the array in whole words (array elements) and $sb = \text{mod}(s, wb)$ the residual rotation in bits (using standard integer division and modulo operations). Clearly $sb < wb$. This division serves the purpose of efficient implementation by a combination of two operations: First the whole array is shifted by $sb$ bits in a way where two successive array elements are combined together into a resulting array element using bitwise operations and where the last array element is also used as the element preceding the first one. Then the rest of the shift of the array by $sw$ words is implemented by reshuffling whole array using modulo indexing. These two steps can be combined together into a single scan of the input array. Algorithm A1 provides the resulting algorithm implementing an arbitrary bit shift on an array where $zw$ is the number of words in array, and $Z = zw \cdot wb$.

Algorithm A1:

```
sW = shift / wb ;                //shift in words
sb = shift % wb ;                //remaining bits
prevSW = A[ ZW - 1 ] ;           //previous word
for( i = 0 ; i < ZW ; i++ ){
    j = ( i + sW ) % ZW ;        //DST word index
    curSW = A[ i ] ;             //source word
    hi = prevSW << wb - sb ;
    lo = curSW >> sb ;
    B[ j ] = hi | lo ;            //bitwise OR
    prevSW = A[ i ] ;
}
```

This algorithm is effective and works for any shift size $s = 1$ to $Z – 1$. With the important limitation of the block size $Z$ equal to a multiple of the word size $wb$. This algorithm will therefore be directly used for encoding half of the supported codeword sizes defined by the WIMAX standard, while being completely unsuitable for Wi-Fi 6.

# 4 Universal LDPC Decoder

In our LDPC decoder implementation, the algorithm choice was motivated by our efforts to produce an ideally exact, or at least practically close approximation of the decoding results obtained when using the widely popular but proprietary MATLAB decoder provided by the `ldpcDecode()` function of the MATLAB R2021b Communications Toolbox . This determined the algorithm choice to be the time proven min-sum algorithm defined in [40,41], which provides high throughput and is a good approximation to the much more computationally demanding posterior Log-Likelihood Ratio (LLR)-based Belief Propagation algorithm [42]. To save memory, the single-scan version of the min-sum algorithm, as described in [43] was implemented, more specifically the layered version with faster convergence as described in [44]. Two versions of the decoder were implemented: one using floating point arithmetic – the 32-bit `float` C type, and another, slightly faster and more memory efficient, fixed point arithmetic decoder, where the LLR approximations were stored as 16-bit `short int` values.

The paper by Huang [43] provides a concise overview of the min-sum algorithm, together with detailed description of its memory optimized single-scan improvement. Both algorithm variants are briefly summed up here: For a Gaussian channel and binary (BPSK) modulation, let $x_n$ be the n-th transmitted bit, $n_n$ the sample of Additive White Gaussian Noise (AWGN), and $y_n$ the sample in the receiver. The modulation and channel transfer can be then described by (16) and the soft-decoding min-sum algorithm works with the LLR metric $Z_n$, defined by (17):

$$y_n = (-2x_n + 1) + n_n \tag{16}$$

$$Z_n^{(0)} = \ln \frac{p(x_n = 0 / y_n)}{p(x_n = 1 / y_n)} = 2 y_n / \sigma^2 \tag{17}$$

where the $\sigma^2$ is the channel variance (may be omitted in the min-sum implementation) and the superscript $(k)$ indicates the iteration number. Each of the iteration implements the following steps:

1. The horizontal scan – the check node update:
   For each *m* and each *N(m)* calculate the check node message $L_{mn}$, based on the variable node messages $Z_{mn}$ coming from all incident variable nodes except the output one (with index *n*).

$$L_{mn}^{(k)} = \min_{n' \in N(m) \setminus n} \left| Z_{mn'}^{(k)} \right| \cdot \prod_{n' \in N(m) \setminus n} \mathrm{sgn}\left(Z_{mn'}^{(k)}\right) \tag{18}$$

2. The vertical scan - the variable node update:
   For each *n* and each *M(n)* calculate the variable node message $Z_{mn}$ coming from all incident check nodes, except the output one (with index *m*).

$$Z_{mn}^{(k)} = Z_n^{(0)} + \sum_{m' \in M(n) \setminus m} L_{m'n}^{(k)} \tag{19}$$

3. For each *n* and each *M(n)* calculate the posterior LLR estimate $Z_n^{(k)}$:

$$Z_n^{(k)} = Z_n^{(0)} + \sum_{m \in M(n)} L_{mn}^{(k)} \tag{20}$$

4. Hard decision and termination:
For each codeword position *n* calculate the bit estimate (21) and if the orthogonality with the **H** matrix condition is satisfied: $\mathbf{H\hat{x}} = \mathbf{0}$, terminate the decoding.

$$\hat{x}_n^{(k)} = \begin{cases} 0, & \text{if } Z_n^{(k)} > 0; \\ 1, & \text{otherwise} \end{cases} \quad (21)$$

The values of $Z_{mn}$ are initialized by received channel samples: $Z_{mn}^{(0)} := Z_n^{(0)}$ and (19) can be easily rewritten using (20):

$$Z_{mn}^{(k)} = Z_n^{(k)} - L_{mn}^{(k)} \quad (22)$$

The single-scan min-sum algorithm modification takes advantage of the following observation: when eq. (18) is applied to a vector of values, the resulting vector only contains elements with two possible magnitudes: the minimum of the input elements, and the second minimum. If signs are stored separately from magnitudes, compressed as a bitmap in a single word, this property, exploited by the algorithm design in [43], can be used to drastically reduce the decoder memory requirements. For example, the maximum check node degree for the Wi-Fi 6 code $R = 5/6$ is $\max(g(m)) = 20$, so this optimization provides a saving of 18 magnitudes for each check node. Furthermore, as described in [43], only the check node messages $L_{mn}$ need to be stored, immediately lowering memory requirements by almost 50%. Equation (18) can combined with (22) and rewritten as:

$$L_{mn}^{(k)} = \min_{n' \in N(m) \setminus n} \left| Z_n^{(k-1)} - L_{mn'}^{(k-1)} \right| \cdot \prod_{n' \in N(m) \setminus n} \text{sgn}(Z_n^{(k-1)} - L_{mn'}^{(k-1)}) \quad (23)$$

To reduce the number of iterations necessary for obtaining the desired BER, a further modification of the computation – the layered decoding described in [44] was implemented. Without an in-depth description we can summarize it in the following way: The horizontal scan in (23) is not performed on each of the *M* check nodes in a single loop, but is organized in tiers - groups of check nodes of size *Z* block-rows of matrix **H**. The calculations of (20), (22) and (23) are then interleaved during each decoder iteration (*k*): (23) is evaluated for a block of *Z* rows/checks, after which the updated values $Z_n$ are used in evaluation of (20) and (22) which in turn updates the inputs for the check nodes calculation (23) of the next tier. The algorithm then consists of two embedded iterations: the inner one alternating between the horizontal and vertical step, and the outer iteration over the $M_b$ groups of check nodes/rows. These are sometimes denoted as super-iterations.

## 5 Fixed Point Implementation

For potential use of the LDPC decoder in constrained systems, especially with memory saving in mind, we also implemented a fixed-point arithmetic decoder, by default using the `int16_t` exact-width integer data type, introduced in C99 for the min-sum algorithm metrics. This usually but not necessarily compiles to `short int`. Because the decoding algorithm only uses the minimum, plus and minus operations, the fixed-point arithmetic can further be simplified by simply mapping of floating point values to an integer range. Our choice of the 16-bit representation is much wider than the various 8- and even fewer bits used in decoders described in literature, but the implementation is fully parametrized, so it is in theory easy to switch to a narrower 8bit representation. It is important to note that, when using narrow integer types, MATLAB uses a saturating arithmetic to minimize errors coming from signal values higher than the maximum level, while the standard C arithmetic uses the faster native wrap-around (or overflow) arithmetic.

Since it would be costly to check for overflow every time an addition is made, the primary way used in our implementation is to use a wider storage than necessary. In the case of the 16bit short `int16_t` C type, the floating point LLR values are actually sampled as $Q_B$ bit integers, with the default value of $Q_B = 10$. With an extra sign bit, this effectively defines an 11bit signed range from -1023 to 1023. These values are stored within 16bit integers, leaving the 5 most significant bits free so when values accumulate during the min-sum algorithm operation, the overflow occurrences will be reasonably rare and not catastrophically effect the resulting BER. Logically we can expect this negative effect to be stronger with a growing number of summation elements, therefore codes with smaller rate should be affected more. This is confirmed in the results section.

In order to support memory constrained architectures, three different configurations with different memory requirements can be defined: Config A – a completely universal and memory unconstrained implementation, will be useful for MATLAB simulations. It supports all of the LDPC code parameters defined in (Tab.1) and (Tab.2) with an array encoder and floating point decoder. Config B present a memory-constrained implementation: a bitmap encoder and fixed-point decoder. It supports only a subset of compatible code (*N*, *K*) parameters, more specifically the half of the WiMAX code parameters, shown in bold in (Tab.2). Config C will be a minimal memory implementation with only one of the LDPC code parameters: *N* = 576 and *R* = 1/2 supported. (Tab.3) gives an overview of the memory requirements for all three implementations:

Table 3. Rounded memory requirements for various configurations in Bytes.

| Configuration | ENC [B] | DEC [B] | Total [KiB rounded] |
|---|---|---|---|
| Config A | 4896 | 162 432 | 164 |
| Config B | 864 | 84 096 | 83 |
| Config C | 240 | 10 080 | 11 |

An important aspect of our implementation is the seamless integration with MATLAB. While writing whole simulations in C is possible, and may provide very high throughput, the flexibility of MATLAB as a tool for rapid prototyping and testing of potential modifications to existing algorithms makes it an indispensable platform. Out of all the implementations available online [28] - [32] none provides both C-language implementation with its performance, and also MATLAB integration. The user can, however, sometimes choose between one of the two. Our implementation fills this niche in that it implements a C encoder and decoder that can be directly used in an actual communication system, while also allowing to run them unchanged from within the MATLAB environment. To achieve this, separate C source files implement the necessary MEX wrappers that make the encoder and decoder functions callable from MATLAB. All code is written in the C99 dialect supporting the portable representation of exact width integer types and tested with the GNU `c99` command, along with `g++`. If necessary, the encapsulation of functions to C++ objects should be a simple task.

For both encoder and decoder the necessary buffers are defined as statically allocated arrays and the important code parameters are written as constants (preprocessor macros) in an automatically generated header file `ldpc.h` and source file `ldpc.c`. This enables for a simple and readable code, where the analogy with the underlying mathematical equations is clearly visible. Furthermore, this design facilitates the shift of C code optimizations to the compiler, where they belong. The compilation to MEX file is done from inside the running MATLAB environment, and appropriate functions are provided to make this process completely transparent to the user, i. e. the user only has to call the provided functions, and doesn't need to care about the details of compilation.

Since many parameters are compiled in, there are some limitations to current usage: the adaptive change of code parameters during the simulation of an Adaptive Coding and Modulation (ACM) systems is not yet implemented, but is still possible. All that needs to be done is to call the compilation function at the beginning of simulation several times, one for each parameter set, and specify slightly different names for the resulting MEX modules. The m-file wrapper would then need to be extended to call a different MEX module for different code parameters. No C code modification would be needed for this extension

## 6 BER and Throughput Analysis

### 6.1 BER Evaluation

We evaluated the error correcting capabilities of our implementation for the QC-LDPC codes used in IEEE 802.11ax and IEEE 802.16-2017 communication standards using the classic approach of Monte-Carlo simulation producing the waterfall curve for the AWGN channel. The whole communication chain: LDPC encoder, channel and LDPC decoder was evaluated alongside the commercial Communication toolbox LDPC implementation introduced in recent MATLAB version R2021b (denoted further as COM). As shown in (Fig.2) with curves for selected code parameters, the error performance of our floating point implementation is visually indistinguishable from the toolbox implementation.

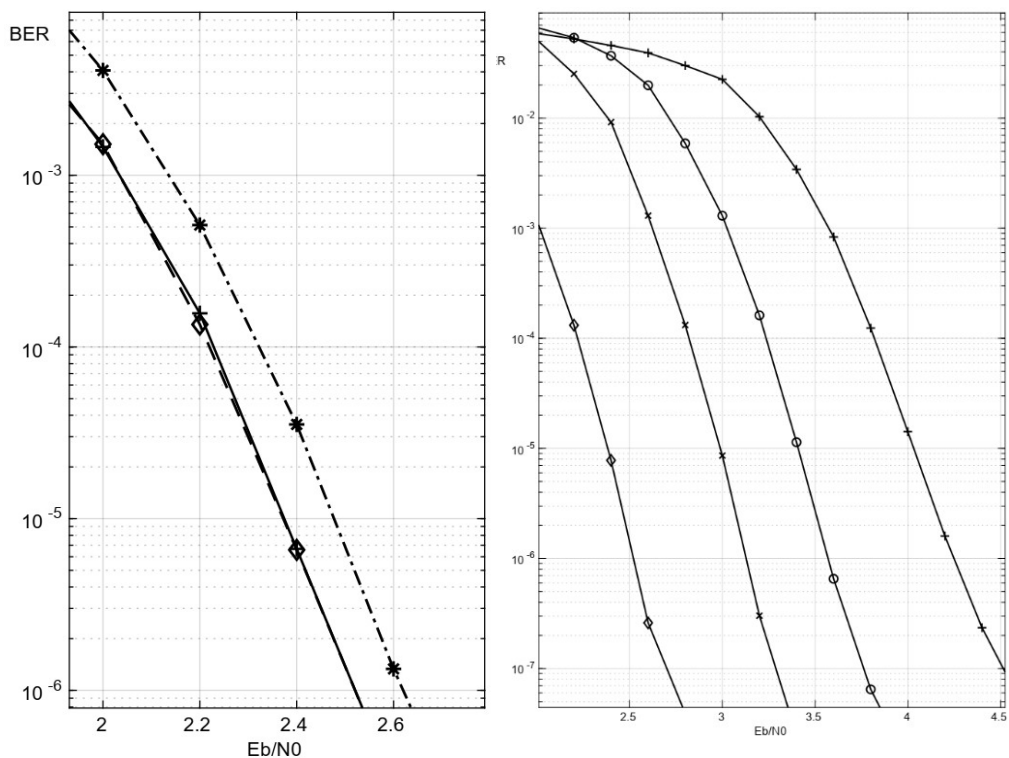

Fig.2. **a) (left)** Error performance for the Wi-Fi 6 R = 1/2 LDPC N = 1944 for maximum 10 decoder iterations along with the MATLAB Communications toolbox R2021b LDPC implementation (plus marker). Our floating point (diamond) result closely copies the COM implementation, fixed point decoder (star) shows some expected penalty. **b) (right)** Results for the IEEE 802.11-2020 defined LDPC codes for N = 1944 and maximum of 10 decoder iterations, Code rates R = 1/2 (diamond), R = 2/3 (x), R = 3/4 (o) and R = 5/6 (+).

As expected, the memory-constrained fixed point decoder (Config B) introduces an approximately 0,1 dB penalty. This effect is most visible for larger values of $N$ and smaller code rates where the eq. (22) sums over large number of parity-check matrix rows; 972 rows for the code with results shown in (Fig.2). For smaller $N = 576$ and higher $R = 5/6$, the fixed-point decoder curve is getting even closer to the floating point implementation. During all the simulations at least 10000 errors were collected for each data point on the waterfall curve.

(Fig.2b) provides a comparison of results for various code rates defined for the $N = 1944$ LDPC code used in Wi-Fi 6, and (Tab.4) gives another perspective on the similarities of our implementation (MEX) and the existing closed-source implementation (COM). Along with the number of simulated bit transfers a direct comparison of the resulting absolute number of errors after decoding is given. More important relative error between the COM and MEX implementation is calculated (denoted $\Delta$).

Table 4. The average number of iterations (nIter), number of decoded bits (in thousands), absolute number of errors of the commercial decoder (#E-COM) and our open source implementation (#E-MEX).

| Eb/N0 | nIter | Kbit | #E-COM | #E-MEX | Δ% |
|---|---|---|---|---|---|
| 1.00 | 8.00 | 1152 | 131 877 | 131 873 | 3.03e-3 |
| 1.20 | 7.99 | 3456 | 313 767 | 313 768 | 3.19e-4 |
| 1.40 | 7.93 | 5760 | 346 651 | 346 651 | 0.0 |
| 1.60 | 7.73 | 8064 | 241 298 | 241 300 | 8.29e-4 |
| 1.80 | 7.17 | 10368 | 97 798 | 97 794 | 4.10e-3 |
| 2.00 | 6.36 | 12672 | 23 731 | 23 732 | 4.20e-3 |

As shown in the last column of (Tab.4), in terms of number of errors after decoding, the implementations give practically the same results. This is also true for the average number of decoder iterations, denoted nIter, so only the value for the MEX decoder is shown.

### 6.2 Throughput Evaluation

Since the LDPC decoder is by far the most computationally intensive task in our simulations, throughput evaluation was focused on the decoder function. Several different single- and multi-threaded decoder configurations were evaluated and throughput compared to the optimized MATLAB R2021b `ldpcDecode()` implementation (denoted as COM). Two x86-64 platforms were compared: MATLAB R2021b on Ubuntu 18.04LTS running on an 8 core/16 thread Intel Core i7-9800X CPU at 3.80GHz with Skylake-X architecture supporting the AVX-512 instruction set extension, and MATLABR2022a on Ubuntu 20.04LTS running on an 16 core/32 thread AMD Ryzen 9 5950X Processor.

(Tab.5) summarizes the data-bits throughput of the decoder for the WIMAX QC-LDPC code with rate $R = 5/6$, and codeword size $N = 2304$. Data were processed in blocks of 10 codewords per thread and the decoder was always set to perform a fixed number of 10 iterations in order to prevent comparing runtimes with varying iteration numbers. 32-bit floating point and 16-bit fixed point (integer) implementations were evaluated along with two different multithreading approaches: The first, more straightforward, implementation spawns worker threads each time the MEX function is run by MATLAB and destroys them during the same call, just after the block is decoded. This is denoted in (Tab.5) with the suffix: simple. A more sophisticated method, denoted as MTX, starts the worker threads once and then synchronizes to them each time the decoder MEX function is called by using POSIX thread conditional variables and mutexes.

Table 5. Throughput comparison for two CPUs: AMD and Intel, single- and multi-threaded implementation, floating- and fixed- point decoder, Commercial LDPC implementation (COM), and our implementation (MEX). MTX denotes a more sophisticated thread synchronization implementation using POSIX mutexes.

|    | LDPC Decoder implementation / CPU | Throughput [Mbps] |
|----|-----------------------------------|-------------------|
| 1  | MEX float, single thread, Intel   | 1.61              |
| 2  | MEX fixed, single thread Intel    | 2.15              |
| 3  | COM single thread, Intel          | 2.62              |
| 4  | MEX float, 16 thread simple, Intel| 11.19             |
| 5  | MEX float, 16 thread MTX, Intel   | 12.40             |
| 6  | MEX fixed, 16 thread simple, Intel| 13.19             |
| 7  | MEX fixed, 16 thread MTX, Intel   | 13.44             |
| 8  | COM Multi thread, Intel           | 2.10              |
| 9  | MEX float, single thread, AMD     | 2.45              |
| 10 | MEX fixed, single thread AMD      | 3.70              |
| 11 | COM single thread, AMD            | 3.64              |
| 12 | MEX float, 32 thread simple, AMD  | 36.46             |
| 13 | MEX float, 32 thread MTX, AMD     | 37.76             |
| 14 | MEX fixed, 32 thread simple, AMD  | 40.35             |
| 15 | MEX fixed, 32 thread MTX, AMD     | 43.95             |
| 16 | COM Multithreaded, AMD            | 3.33              |
| 17 | CLI float, single thread AMD      | 6.31              |

As shown on lines 1 to 3, and 9 to 11 our single-threaded implementation reaches only about 60% of the otherwise equivalent Communications Toolbox function. This is not surprising, given the fact that the LDPC decoder in MATLAB is now a mature optimized implementation, which is only available as a closed-source MEX file. It is hard to infer what optimizations, such as the use of intrinsics or the optimized Intel MKL library, were made. Our implementation relies on the standard C99 language constructs only, which brings some throughput penalty but facilitates portability. The fixed point implementation improves the throughput slightly while sacrificing some error performance.

Since modern CPUs have been using many cores for a long time, the multi-threaded implementation is actually the one that matters. Here the advantage of our approach lies in the ability to fine-tune the number of threads that the user can specify explicitly, compared to the On/Off setting in the toolbox function. Lines 4 to 7 and 12 to 15 compare various multithreaded implementations on Intel and AMD platforms. The comparison with the toolbox function is given on lines 8 and 16, showing the throughput of the multithreaded version of the `ldpcDecode()` method to be an order of magnitude lower than of our implementation. The low throughput of the multithreaded toolbox decoder is somewhat surprising, and may indicate a bug in the implementation, potentially fixed in some later toolbox release. The actual use of multiple threads was checked by the OS-built-in system monitor utility.

What's a bit disappointing is the almost negligible throughput improvement of the more sophisticated MTX design, where the worker threads are running (or waiting) in parallel to the main MATLAB thread, and are synchronized to the main thread by means of POSIX conditional variables and mutexes. As shown in rows 5, 7, 13, 15 of (Tab.5), these actually represent so much overhead that such implementation, with its greatly complicated design, bring negligible benefit relative to the simple implementation shown in rows 4, 6, 12, and 14.

For more insight into how much the MATLAB environment affects performance, a command line version of the benchmark was run, compiled by the GNU `c99` CLI compiler. The result shown on line 17 of (Tab.5) indicate more than double the performance.

*High Throughput Open-Source Implementation of Wi-Fi 6 and WiMAX LDPC Encoder-Decoder*# Conclusion

In this paper we described the details of our QC-LDPC encoder and universal LDPC decoder C99 implementations, which focus on (but are not limited to) the modern LDPC codes defined in the IEEE 802.11-2020 and IEEE 802.16-2017 standards. We provided error performance evaluation results comparable to a state-of-the art closed-source implementation provided by the MATLAB R2021b Communications Toolbox, along with comparison of throughput of various configurations on Intel and AMD platforms. The complete source code of the C99 encoder and decoder, along with MATLAB MEX wrappers and supporting MATLAB scripts are freely available at our GitHub page [46] published under a permissive BSD license.

# Acknowledgement

This work was supported by the Slovak APVV Agency under cont. no. APVV-19-0436.# References

[1]   TANNER, R. M., SRIDHARA, D., SRIDHARAN, A., et al. LDPC block and convolutional codes based on circulant matrices. IEEE Transactions on Information Theory, 2004, vol. 50, no. 12, pp. 2966-2984. DOI: 10.1109/TIT.2004.838370

[2]   LI, Z., CHEN, L., ZENG, L., et al. Efficient encoding of quasi-cyclic low-density parity-check codes. IEEE Transactions on Communications, 2006, vol. 54, no. 1, pp. 71-81. DOI:10.1109/TCOMM.2005.861667

[3]   IEEE Std. 802.16-2017, IEEE standard for air interface for broadband wireless access systems – Section 8.4.9.2.5: low density parity check (LDPC) code. IEEE New York (USA), 2018, pp. 1459 – 1463. ISBN 978-1-5044-4474-3

[4]   IEEE Std. 802.11ax-2021, Part 11: wireless lan medium access control (MAC) and physical layer (PHY) specifications - Amendment 1: enhancements for high-efficiency WLAN. IEEE New York (USA), 2021. ISBN 978-1-5044-7389-7

[5]   IEEE Std 802.11-2020, Part 11: wireless lan MAC and PHY specifications - Annex F: HT LDPC matrix definitions. IEEE New York (USA), 2021. p. 4130–4132. ISBN 978-1-5044-7283-8

[6]   ETSI EN 302 307 V1.2.1, Digital Video Broadcasting (DVB): Second generation framing structure, channel coding and modulation systems for broadcasting, interactive services, news gathering and other broadband satellite applications (DVB-S2). ETSI Sophia Antipolis Cedex (France), 2009.

[7]   3GPP TS 38.212 V17.4.0, 3rd Generation Partnership Project; Technical Specification Group Radio Access Network; NR; Multiplexing and channel coding (Release 17). 3GPP Valbonne (France), 2022. Available at: http://www.3gpp.org

[8]   IEEE Std. 802.3-2018, IEEE Standard for Ethernet 802.3-Section 7: 10G Ethernet, Chapter: 101.3.2.4 low density parity check (LDPC) forward error correction (FEC) codes. IEEE New York (USA), 2018, p. 321–330. DOI: 10.1109/IEEESTD.2018.8457469

[9]   IEEE Std. 802.3-2018, IEEE Standard for Ethernet 802.3-2018 (Revision of IEEE Std 802.3-2015) - Section 8: 200G and 400G Ethernet Chapter: 119.2.4.6. IEEE New York (USA), 2018, p.67–69. DOI: 10.1109/IEEESTD.2018.8457469

[10]  FARKAS, P., RAKUS, M. Decoding five times extended reed solomon codes using syndromes. Computing and Informatics, 2020, no. 6, vol 39, pp.1311–1335. ISSN 2585-8807 (online), DOI: 10.31577/cai_2020_6_131129

## ◤ Authors

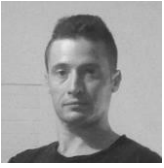

**Ing. Tomáš Páleník, PhD.**
Slovak University of Technology in Bratislava, Slovakia
tomas.palenik@stuba.sk (corresponding author)
He received his Master's degree in 2006 from the STU and in 2010 he finalized his dissertation: "Communication system design based on an SDR platform: Exploiting the redundancy of an OFDM system" and received the Ph.D. degree in telecommunications. His research interests include digital communications systems, Orthogonal Frequency Division Multiplexing, Software Defined Radio, Error-Control Coding, IPv6 & IoT and graph algorithms in communications. During 2006–2008 he worked as an external consultant for Sandbridge-Technologies, N.Y., USA, where he implemented an LDPC decoder for an in-house developed multicore mobile processor SandBlaster. Later working in academia, he has participated in multiple research projects and also took part in the COST action IC1407. In 2019 he worked as a researcher and analyst at the AIT - Austrian Institute of Technology. He presented at several international conferences such as Wireless Innovation's SDR in Washington D.C., USA.
He is a member of the IEEE

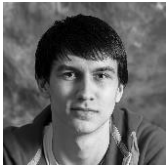

**Bc. Viktor Szitkey**
Slovak University of Technology in Bratislava, Slovakia
xszitkey@stuba.sk
Research interests focuses on various technology implementation in SDR.
Student member of the IEEE.